\documentclass[proceedings]{JHEP3}
\usepackage{amsfonts}
\usepackage{amsmath}
\usepackage{epsfig,multicol}

\newbox\mybox

\newcommand\fverb{\setbox\mybox=\hbox\bgroup\verb}
\newcommand\fverbdo{\egroup\medskip\noindent\fbox{\unhbox\mybox}\ }
\newcommand\fverbit{\egroup\item[\fbox{\unhbox\mybox}]}
\conference{PT-symmetric noncommutative spaces}
\abstract{We provide a systematic procedure to relate a three dimensional q-deformed oscillator algebra to the corresponding algebra satisfied by canonical variables describing noncommutative spaces. The large number of possible free parameters in these calculations is reduced to a manageable amount by imposing various different versions of $\mathcal{PT}$-symmetry on the underlying spaces, which are dictated by the specific physical problem under consideration. The representations for the corresponding operators are in general non-Hermitian with regard to standard inner products and obey algebras whose uncertainty relations lead to minimal length, areas or volumes in phase space. We analyze in particular one three dimensional solution
which may be decomposed to a two dimensional noncommutative space plus one commuting space component and also into a one dimensional noncommutative space plus two commuting space components. We study
some explicit models on these type of noncommutative spaces.}

\title{$\mathcal{PT}$-symmetric noncommutative spaces with minimal volume
uncertainty relations}
\author{Sanjib Dey$^\bullet$, Andreas Fring$^\bullet$ and Laure Gouba$^\circ 
$ \\
$^\bullet$ Centre for Mathematical Science, City University London,\\
$\,\,$ Northampton Square, London EC1V 0HB, UK\\
$^\circ$ The Abdus Salam International Centre for Theoretical Physics
(ICTP), \\
$\,\,$ Strada Costiera 11, 34014, Trieste, Italy \\
E-mail: sanjib.dey.1@city.ac.uk, a.fring@city.ac.uk, lgouba@ictp.it}

\input{tcilatex}

\begin{document}

\section{Introduction}

The simplest and most commonly studied version of noncommutative spaces
replaces the standard set of commuting coordinates by new ones obeying $%
[x^{\mu },x^{\nu }]=i\theta ^{\mu \nu }$, with $\theta ^{\mu \nu }$ being a
constant antisymmetric tensor. However, even in the very first proposals on
noncommutative spaces \cite{Snyder:1946qz} the tensor $\theta ^{\mu \nu }$
was taken to be a function of the position coordinates, i.e. $\theta ^{\mu
\nu }(x)$. Further possibilities arise when one breaks the Lorentz
invariance of the tensor and allows for a general dependence of position and
momenta \cite{Kempf1,Kempf2,Gomes:2009tk,Gomes:2009rz}. It is known for some
time that such a scenario leads to the interesting versions of a generalized
version of Heisenberg's uncertainty relations \cite{Kempf1,Kempf2}. In
particular when the commutation relations are modified in such a way that
their structure constants involve higher powers of the momenta or
coordinates one encounters minimal lengths or momenta, respectively. As a
consequence of these type of relations lead to more radical changes in the
interpretation of possible measurements than the conclusions usually drawn
from the standard relations. Whereas the conventional relations, which
simply have Planck's constant $\hbar $ as structure constant, only prevent
that two quantities commuting in this manner, e.g. $x$ and $p$, can be known
simultaneously, the modified versions prohibit that the observables can be
known at all below a certain scale, the minimal length or minimal momentum.
This scale is usually identified to be of the order of the Planck scale.
Combining some of these minimal lengths in two dimension leads to minimal
areas and in three dimensions to minimal volumes. The need for such type of
noncommutative space structures has arisen in many contexts, such as for
instance in certain string theories \cite{wit} and models investigating
gravitational stability \cite{dop}. For a general review on noncommutative
quantum mechanics see for instance \cite{delduc} and for a review on
noncommutative quantum field theories see \cite{Douglas:2001ba,Szabo:2001kg}.

By now many studies on the structure of such type of generalized canonical
relations have been carried out \cite%
{Kempf1,Kempf2,Brodimas,Bieden,MacF,ChangPu,QuesneTK,AFBB,Hossenfelder,AFLGFGS,AFLGBB,Chang:2011jj,Lewis:2011fg,Pedram:2011xj,Dorsch:2011qf,Nozari:2012gd,Gavrilik:2012yj}%
, albeit mostly in dimensions less than three. Besides leading to different
physical results, a furher crucial difference between the Snyder type
noncommutative spaces and those with broken Lorentz invariance is the way
they are constructed. Whereas the former spaces can be thought of as arising
naturally from deformations based on general twists \cite%
{Aschieri:2005zs,Castro:2011in}, the construction of the latter is less
systematic and is usually based on the deformation of oscillator algebras 
\cite{Schwenk:1992sq,Brodimas,Kempf2}. In \cite{AFBB,AFLGFGS,AFLGBB} it was
shown in one and two dimensions, respectively, how to map q-deformed
oscillator algebras onto canonical variables. The main purpose of this
manuscript is to extend these considerations to the full three dimensional
space. This approach has the advantage that it allows for the explicit
construction of the entire Fock space \cite{Bieden,MacF,ChangPu}.

Our manuscript is organized as follows: As an introduction we explain in
section 2 how one can systematically construct canonical variables on a flat
noncommutative space starting from standard generators in Fock space by
exploiting $\mathcal{PT}$-symmetry to reduce the number of free parameters.
In section 3 we extend these considerations to q-deformed oscillator
algebras and present in particular one solution in more detail for which we
construct in a nontrivial limit a non-Hermitian representation. In section 4
we demonstrate that, depending on the dimension, this solution leads to
minimal length, areas or volumes. In section 5 we study the non-Hermitian $%
\mathcal{PT}$-symmetric versions of the harmonic oscillator on these spaces
and in section 6 we state our conclusions.

\section{Deformed oscillator algebras and noncommutative spaces}

Our starting point is a $q$-deformed oscillator algebra for the creation and
annihilation operators $A_{i}^{\dagger }$, $A_{i}$ as studied for instance
in \cite{Bieden,MacF,ChangPu,AFBB,AFLGBB} 
\begin{equation}
A_{i}A_{j}^{\dagger }-q^{2\delta _{ij}}A_{j}^{\dagger }A_{i}=\delta
_{ij},\quad \lbrack A_{i}^{\dagger },A_{j}^{\dagger }]=0,\quad \lbrack
A_{i},A_{j}]=0,\qquad \text{for }i,j=1,2,3;q\in \mathbb{R}.  \label{AAAA}
\end{equation}%
In the limit $q\rightarrow 1$ we denote $A_{i}\rightarrow a_{i}$ and recover
the standard Fock space commutation relations 
\begin{equation}
\lbrack a_{i},a_{j}^{\dagger }]=\delta _{ij},\qquad \lbrack
a_{i},a_{j}]=0,\qquad \lbrack a_{i}^{\dagger },a_{j}^{\dagger }]=0,\qquad 
\text{for }i,j=1,2,3.  \label{Fock}
\end{equation}%
We assume further that the relations (\ref{Fock}) are linearly related to
the standard three dimensional flat noncommutative space characterized by
the relations%
\begin{equation}
\begin{array}{lll}
\lbrack x_{0},y_{0}]=i\theta _{1}, & [x_{0},z_{0}]=i\theta _{2},\qquad \quad
& [y_{0},z_{0}]=i\theta _{3}, \\ 
\lbrack x_{0},p_{x_{0}}]=i\hbar ,\qquad \quad & [y_{0},p_{y_{0}}]=i\hbar , & 
[z_{0},p_{z_{0}}]=i\hbar ,%
\end{array}%
~~\ \ \ \ \ \text{for }\theta _{1},\theta _{2},\theta _{3}\in \mathbb{R},
\label{cano}
\end{equation}%
with all remaining commutators to be zero and the $\theta _{1},\theta
_{2},\theta _{3}$ denoting the noncommutative constants. The most general
linear Ansatz to relate the generators of relations (\ref{cano}) and (\ref%
{Fock}) reads 
\begin{equation}
\varphi _{i}=\sum\limits_{j=1}^{3}\kappa _{ij}~a_{j}+\lambda
_{ij}a_{j}^{\dagger },~~\ \ \ \ \ \ \text{for }\vec{\varphi}%
=\{x_{0},y_{0},z_{0},p_{x_{0}},p_{y_{0}},p_{z_{0}}\},  \label{sd}
\end{equation}%
where the $\kappa _{ij}$, $\lambda _{ij}$ having dimensions of length or
momentum for $i=1,2,3$ or $i=4,5,6$ respectively. The commutation relations
obeyed by the canonical variables $X$, $Y$, $Z$, $P_{x}$, $P_{y}$, $P_{z}$
associated to the deformed algebra (\ref{AAAA}) are yet unknown and are
subject to construction. The algebra they satisfy may be related to (\ref%
{AAAA}) by similar relations as (\ref{sd}), but since the constants $\kappa
_{ij}$ and $\lambda _{ij}$ are in general complex, this amounts to finding $%
72$ real parameters. To reduce this number to a manageable quantity one can
utilize $\mathcal{PT}$-symmetry.

\subsection{The role of $\mathcal{PT}$-symmetry}

Whereas the momenta and coordinates in (\ref{cano}) are Hermitian operators
acting on a Hilbert space with standard inner product, this is no longer
true for the variables associated to the deformed algebra (\ref{AAAA}) as
they become in general non-Hermitian with regard to these inner products.
Thus a quantum mechanical or quantum field theoretical models on these
spaces will, in general, not be Hermitian in that space. However, it is by
now well accepted that one may consider complex $\mathcal{PT}$ symmetric
non-Hermitian systems as self-consistent descriptions of physical systems 
\cite{Bender:1998ke,Benderrev}. Guided by these results one may try to
identify this symmetry for the noncommutative space relations (\ref{cano}).
In \cite{Giri:2008iq} the authors argue that this would not be possible and
one is therefore forced to take the noncommutative constants to be complex.
We reason here that this is incorrect and even the standard noncommutative
space relations are in fact symmetric under many different versions of $%
\mathcal{PT}$-symmetry. All one requires to formulate a consistent quantum
description is an antilinear involutory map \cite{EW} that leaves the
relations (\ref{cano}) invariant. We identify here the several possibilities:

Taking for instance $\theta _{2}=0$, the algebra (\ref{cano}) remains
invariant under the following antilinear transformations 
\begin{equation}
\begin{array}{cllll}
\mathcal{PT}_{\pm }:\quad & x_{0}\rightarrow \pm x_{0}, & y_{0}\rightarrow
\mp y_{0}, & z_{0}\rightarrow \pm z_{0}, & i\rightarrow -i, \\ 
& p_{x_{0}}\rightarrow \mp p_{x_{0}}, & p_{y_{0}}\rightarrow \pm p_{y_{0}},
& p_{z_{0}}\rightarrow \mp p_{z_{0}}. & 
\end{array}
\label{+-}
\end{equation}%
We may also attempt to keep $\theta _{2}$ different from zero, in which case
we have to transform the $\theta _{2}$ as well in order to achieve the
invariance of (\ref{cano}) 
\begin{equation}
\begin{array}{cllll}
\mathcal{PT}_{\mathcal{\theta }_{\pm }}:\quad & x_{0}\rightarrow \pm x_{0},
& y_{0}\rightarrow \mp y_{0}, & z_{0}\rightarrow \pm z_{0}, & i\rightarrow
-i, \\ 
& p_{x_{0}}\rightarrow \mp p_{x_{0}}, & p_{y_{0}}\rightarrow \pm p_{y_{0}},
& p_{z_{0}}\rightarrow \mp p_{z_{0}}, & \theta _{2}\rightarrow -\theta _{2}.%
\end{array}%
\end{equation}%
A further option, which also allows to keep $\theta _{2}$ different from
zero, would be to introduce permutations amongst the different directions%
\begin{equation}
\begin{array}{cllll}
\mathcal{PT}_{xz}:\quad & x_{0}\rightarrow z_{0}, & y_{0}\rightarrow y_{0},
& z_{0}\rightarrow x_{0}, & i\rightarrow -i, \\ 
& p_{x_{0}}\rightarrow -p_{z_{0}}, & p_{y_{0}}\rightarrow -p_{y_{0}}, & 
p_{z_{0}}\rightarrow -p_{x_{0}}. & 
\end{array}
\label{xz}
\end{equation}%
Clearly all of these maps are involutions $\mathcal{PT}^{2}=\mathbb{I}$. In
fact, there might in fact be more options. The occurrence of various
possibilities to implement the antilinear symmetry is a known feature
previously observed for many examples \cite{chainOla,Mon1,Paulos} in
dimensions larger than one. More restrictions and the explicit choice of
symmetry result from the specific physical situation one wishes to describe.
For instance $\mathcal{PT}_{\pm }$ might be appropriate when one deals with
a problem in which one direction is singled out, $\mathcal{PT}_{\mathcal{%
\theta }_{\pm }}$ requires the noncommutative constant $\theta _{2}$ to
appear as a parameter in the model and $\mathcal{PT}_{xz}$ suggest a
symmetry along the line $x_{0}=z_{0}$. For the creation and annihilation
operators this symmetry could manifest itself in different ways, for
instance as $a_{i}\rightarrow \pm a_{i}$, $a_{i}^{\dagger }\rightarrow \pm
a_{i}^{\dagger }$ or by the permutation of indices $a_{i}\rightarrow
a_{j}^{\dagger }$, $a_{i}^{\dagger }\rightarrow a_{j}$ when they label for
instance particles in different potentials, see e.g. \cite{Eva}. Once again
the underlying physics will dictate which version one should select. The
general reason for the occurrence of these different possibilities are just
manifestations of the ambiguities in defining a metric to which the $%
\mathcal{PT}$-operator is directly related. What needs to be kept in mind is
that we only require the symmetry of some antilinear involution \cite{EW} in
order to obtain a meaningful quantum mechanical description.

\subsection{Oscillator algebras of flat noncommutative spaces}

Let us first see how to represent a three dimensional oscillator algebra in
terms of the canonical variables in three dimensional flat noncommutative
space. For definiteness we seek at first a description which is invariant
under $\mathcal{PT}_{\pm }$. The most generic linear Ansatz for the creation
and annihilation operators to achieve this is 
\begin{eqnarray}
a_{1} &=&\alpha _{1}x_{0}+i\alpha _{2}y_{0}+\alpha _{3}z_{0}+i\alpha
_{4}p_{x_{0}}+\alpha _{5}p_{y_{0}}+i\alpha _{6}p_{z_{0}},  \label{a1} \\
a_{2} &=&\alpha _{7}x_{0}+i\alpha _{8}y_{0}+\alpha _{9}z_{0}+i\alpha
_{10}p_{x_{0}}+\alpha _{11}p_{y_{0}}+i\alpha _{12}p_{z_{0}},  \label{a2} \\
a_{3} &=&\alpha _{13}x_{0}+i\alpha _{14}y_{0}+\alpha _{15}z_{0}+i\alpha
_{16}p_{x_{0}}+\alpha _{17}p_{y_{0}}+i\alpha _{18}p_{z_{0}},  \label{a3}
\end{eqnarray}%
with dimensional real constants $\alpha _{i}$. We note that we have $%
\mathcal{PT}_{\pm }:a_{i}\rightarrow \pm a_{i},a_{i}^{\dagger }\rightarrow
\pm a_{i}^{\dagger }$ for $i=1,2,3$. The nonsequential ordering of the
constants in (\ref{a1})-(\ref{a3}) is chosen to perform the limit to the two
dimensional case in a convenient way. For $\alpha _{9},\ldots ,\alpha
_{18}\rightarrow 0$ we recover equation (2.4) in \cite{AFLGBB}. It is useful
to invoke this limit at various stages of the calculation as a consistency
check. We then compute that the operators (\ref{a1})-(\ref{a3}), expressed
on the three dimensional flat noncommutative space (\ref{cano}), satisfy the
standard Fock space commutation relations (\ref{Fock}) provided that the
following nine constraints hold%
\begin{equation}
1=2\sum\limits_{j=1}^{3}\left[ (2-j)\alpha _{2+k}\alpha _{j+k}\theta
_{j}-(-1)^{j}\hbar \alpha _{j+k}\alpha _{j+k+3}\right] \ \qquad \text{for }%
k=0,6,12,  \label{e1}
\end{equation}%
\begin{eqnarray}
0 &=&i(\alpha _{p}\alpha _{q}+\alpha _{p+2}\alpha _{q-2})\theta
_{2}+\sum\limits_{j=1}^{2}(\alpha _{j+p}\alpha _{j+q-p+2}-\alpha
_{j+p-1}\alpha _{j+q-2})\theta _{2j-1} \\
&&+\sum\limits_{j=1}^{3}(\alpha _{j+p+2}\alpha _{j+q-p-2}-\alpha
_{j+p-1}\alpha _{j+q})\hbar ~~~~\ \ \ \ \text{for }\{p,q\}=\{1,9\},\{1,15\},%
\{7,15\},  \notag \\
0 &=&i(\alpha _{p}\alpha _{q}+\alpha _{p+2}\alpha _{q-2})\theta
_{2}-\sum\limits_{j=1}^{2}(-1)^{j}(\alpha _{j+p}\alpha _{j+q-p+2}+\alpha
_{j+p-1}\alpha _{j+q-2})\theta _{2j-1}  \label{e14} \\
&&-\sum\limits_{j=1}^{3}(-1)^{j}(\alpha _{j+p+2}\alpha _{j+q-p-2}+\alpha
_{j+p-1}\alpha _{j+q})\hbar ~~~\text{for }\{p,q\}=\{1,9\},\{1,15\},\{7,15\}.
\notag
\end{eqnarray}

It turns out that when keeping $\theta _{2}\neq 0$ these equations do not
admit a nontrivial solution. However, setting $\theta _{2}$ to zero we can
solve (\ref{e1})-(\ref{e14}) for instance by%
\begin{eqnarray}
\alpha _{2} &=&-\frac{\alpha _{14}\left( \alpha _{7}\left( 2h\alpha
_{14}\alpha _{17}-2\alpha _{13}\Delta ^{\prime }+1\right) -2\alpha
_{9}\alpha _{13}\Delta ^{\prime \prime }\right) }{2\Delta \Delta ^{\prime
\prime }}, \\
\alpha _{4} &=&\frac{h\alpha _{7}\alpha _{16}\left( -2h\alpha _{14}\alpha
_{17}+2\alpha _{13}\Delta ^{\prime }-1\right) +\alpha _{9}\left( 2h\alpha
_{13}\alpha _{16}-1\right) \Delta ^{\prime \prime }}{2h\Delta \Delta
^{\prime \prime }}, \\
\alpha _{5} &=&\frac{\alpha _{1}\Delta ^{\prime }+\alpha _{3}\Delta ^{\prime
\prime }}{h\alpha _{14}}, \\
\alpha _{6} &=&\frac{2h\alpha _{9}\alpha _{13}\alpha _{18}\Delta ^{\prime
\prime }+\alpha _{7}\left( 2h\alpha _{13}\alpha _{18}\Delta ^{\prime
}-\alpha _{14}\left( 2\alpha _{17}\alpha _{18}h^{2}+\theta _{3}\right)
\right) }{2h\Delta \Delta ^{\prime \prime }}, \\
\alpha _{8} &=&\frac{\alpha _{14}\left( \alpha _{1}\left( 2h\alpha
_{14}\alpha _{17}-2\alpha _{13}\Delta ^{\prime }+1\right) +2\alpha
_{3}\alpha _{13}\left( \alpha _{14}\theta _{3}-h\alpha _{18}\right) \right) 
}{2\Delta \Delta ^{\prime \prime }}, \\
\alpha _{10} &=&\frac{h\alpha _{1}\alpha _{16}\left( 2h\alpha _{14}\alpha
_{17}-2\alpha _{13}\Delta ^{\prime }+1\right) -\alpha _{3}\left( 2h\alpha
_{13}\alpha _{16}-1\right) \Delta ^{\prime \prime }}{2h\Delta \Delta
^{\prime \prime }}, \\
\alpha _{11} &=&\frac{\alpha _{7}\Delta ^{\prime }+\alpha _{9}\Delta
^{\prime \prime }}{h\alpha _{14}}, \\
\alpha _{12} &=&\frac{-2h\alpha _{3}\alpha _{13}\alpha _{18}\Delta ^{\prime
\prime }+\alpha _{1}\left( \alpha _{14}\left( 2\alpha _{17}\alpha
_{18}h^{2}+\theta _{3}\right) -2h\alpha _{13}\alpha _{18}\Delta ^{\prime
}\right) }{2h\Delta \Delta ^{\prime \prime }}, \\
\alpha _{15} &=&\frac{2h\alpha _{14}\alpha _{17}-2\alpha \Delta ^{\prime }+1%
}{2\Delta ^{\prime \prime }},
\end{eqnarray}%
where we abbreviated $\Delta :=\alpha _{3}\alpha _{7}-\alpha _{1}\alpha _{9}$%
, $\Delta ^{\prime }:=h\alpha _{16}+\alpha _{14}\theta _{1}$, $\Delta
^{\prime \prime }:=h\alpha _{18}-\alpha _{14}\theta _{3}$. Thus we still
have nine parameters left at our disposal. In other words the Ansatz (\ref%
{a1})-(\ref{a3}) together with (\ref{Fock}) enforces the $\mathcal{PT}$-symmetry of the type (\ref{+-}).

Inverting the relations (\ref{a1})-(\ref{a3}) we may express the dynamical
variables in terms of the creation and annihilation operators%
\begin{eqnarray}
x_{0} &=&\frac{\alpha _{9}\alpha _{17}-\alpha _{11}\alpha _{15}}{2\det M_{1}}%
(a_{1}+a_{1}^{\dagger })+\frac{\alpha _{5}\alpha _{15}-\alpha _{3}\alpha
_{17}}{2\det M_{1}}(a_{2}+a_{2}^{\dagger })+\frac{\alpha _{3}\alpha
_{11}-\alpha _{5}\alpha _{9}}{2\det M_{1}}(a_{3}+a_{3}^{\dagger }),
\label{x1} \\
y_{0} &=&\frac{\alpha _{10}\alpha _{18}-\alpha _{12}\alpha _{16}}{2i\det
M_{2}}(a_{1}-a_{1}^{\dagger })+\frac{\alpha _{6}\alpha _{16}-\alpha
_{4}\alpha _{18}}{2i\det M_{2}}(a_{2}-a_{2}^{\dagger })+\frac{\alpha
_{4}\alpha _{12}-\alpha _{6}\alpha _{10}}{2i\det M_{2}}(a_{3}-a_{3}^{\dagger
}),~~~~~~~ \\
z_{0} &=&\frac{\alpha _{11}\alpha _{13}-\alpha _{7}\alpha _{17}}{2\det M_{1}}%
(a_{1}+a_{1}^{\dagger })+\frac{\alpha _{1}\alpha _{17}-\alpha _{5}\alpha
_{13}}{2\det M_{1}}(a_{2}+a_{2}^{\dagger })+\frac{\alpha _{5}\alpha
_{7}-\alpha _{1}\alpha _{11}}{2\det M_{1}}(a_{3}+a_{3}^{\dagger }),
\end{eqnarray}%
\begin{eqnarray}
p_{x_{0}} &=&\frac{\alpha _{12}\alpha _{14}-\alpha _{8}\alpha _{18}}{2i\det
M_{2}}(a_{1}-a_{1}^{\dagger })+\frac{\alpha _{2}\alpha _{18}-\alpha
_{6}\alpha _{14}}{2i\det M_{2}}(a_{2}-a_{2}^{\dagger })+\frac{\alpha
_{6}\alpha _{8}-\alpha _{2}\alpha _{12}}{2i\det M_{2}}(a_{3}-a_{3}^{\dagger
}),~~~~~~~ \\
p_{y_{0}} &=&\frac{\alpha _{7}\alpha _{15}-\alpha _{9}\alpha _{13}}{2\det
M_{1}}(a_{1}+a_{1}^{\dagger })+\frac{\alpha _{3}\alpha _{13}-\alpha
_{1}\alpha _{15}}{2\det M_{1}}(a_{2}+a_{2}^{\dagger })+\frac{\alpha
_{1}\alpha _{9}-\alpha _{3}\alpha _{7}}{2\det M_{1}}(a_{3}+a_{3}^{\dagger }),
\\
p_{z_{0}} &=&\frac{\alpha _{8}\alpha _{16}-\alpha _{10}\alpha _{14}}{2i\det
M_{2}}(a_{1}-a_{1}^{\dagger })+\frac{\alpha _{4}\alpha _{14}-\alpha
_{2}\alpha _{16}}{2i\det M_{2}}(a_{2}-a_{2}^{\dagger })+\frac{\alpha
_{2}\alpha _{10}-\alpha _{4}\alpha _{8}}{2i\det M_{2}}(a_{3}-a_{3}^{\dagger
}),  \label{x6}
\end{eqnarray}%
where the matrices $M_{1/2}$ have entries%
\begin{equation}
(M_{l})_{jk}=6j+2k+l-8\qquad \text{for\qquad }l=1,2.
\end{equation}%
These expressions satisfy the commutation relations(\ref{cano}) when we
invoke the constraints (\ref{e1})-(\ref{e14}) and the standard Fock space
relation (\ref{Fock}). In that case we also have the simple relation $\det
M_{1}\det M_{2}=-1/8\hbar ^{3}$. By changing the Ansatz (\ref{a1})-(\ref{a3}%
) appropriately one may also obtain $\mathcal{PT}_{\mathcal{\theta }_{\pm }}$
or $\mathcal{PT}_{xz}$-invariant solutions.

\section{Noncommutative space-time from q-deformed creation and annihilation
operators}

Next we construct the commutation relations for the deformed noncommutative
space satisfied by the canonical variables $X$, $Y$, $Z$, $P_{x}$, $P_{y}$, $%
P_{z}$, which we express linearly in terms of the creation and annihilation
operators obeying the deformed algebra (\ref{AAAA}). Guided by the fact that
in the limit $q\rightarrow 1$ we should recover the relations (\ref{x1})-(%
\ref{x6}) of the previous subsection. We therefore make the Ansatz%
\begin{eqnarray}
X &=&\hat{\kappa}_{1}(A_{1}^{\dagger }+A_{1})+\hat{\kappa}%
_{2}(A_{2}^{\dagger }+A_{2})+\hat{\kappa}_{3}(A_{3}^{\dagger }+A_{3}),
\label{an1} \\
Y &=&i\hat{\kappa}_{4}(A_{1}^{\dagger }-A_{1})+i\hat{\kappa}%
_{5}(A_{2}^{\dagger }-A_{2})+i\hat{\kappa}_{6}(A_{3}^{\dagger }-A_{3}), \\
Z &=&\hat{\kappa}_{7}(A_{1}^{\dagger }+A_{1})+\hat{\kappa}%
_{8}(A_{2}^{\dagger }+A_{2})+\hat{\kappa}_{9}(A_{3}^{\dagger }+A_{3}), \\
P_{x} &=&i\check{\kappa}_{10}(A_{1}^{\dagger }-A_{1})+i\check{\kappa}%
_{11}(A_{2}^{\dagger }-A_{2})+i\check{\kappa}_{12}(A_{3}^{\dagger }-A_{3}),
\\
P_{y} &=&\check{\kappa}_{13}(A_{1}^{\dagger }+A_{1})+\check{\kappa}%
_{14}(A_{2}^{\dagger }+A_{2})+\check{\kappa}_{15}(A_{3}^{\dagger }+A_{3}), \\
P_{z} &=&i\check{\kappa}_{16}(A_{1}^{\dagger }-A_{1})+i\check{\kappa}%
_{17}(A_{2}^{\dagger }-A_{2})+i\check{\kappa}_{18}(A_{3}^{\dagger }-A_{3}),
\label{an6}
\end{eqnarray}%
with $\hat{\kappa}_{i}=\kappa _{i}\sqrt{\hbar /(m\omega )}$ for $i=1,\ldots
,9$ having the dimension of a length and $\check{\kappa}_{i}=\kappa _{i}%
\sqrt{m\omega \hbar }$ for $i=10,\ldots ,18$ possessing the dimension of a
momentum. The constants $\kappa _{i}$ for $i=1,\ldots ,18$ are therefore
dimensionless. We deliberately keep here all dimensional constants different
from $1$. With the help of the $q$-deformed oscillator algebra (\ref{AAAA})
we compute%
\begin{eqnarray}
\lbrack X,Y] &=&2i\sum\nolimits_{j=1}^{3}\hat{\kappa}_{j}\hat{\kappa}_{3+j}%
\left[ 1+\left( q^{2}-1\right) \right] A_{j}^{\dagger }A_{j},  \label{sdd} \\
\lbrack Y,Z] &=&-2i\sum\nolimits_{j=1}^{3}\hat{\kappa}_{3+j}\hat{\kappa}%
_{6+j}\left[ 1+\left( q^{2}-1\right) \right] A_{j}^{\dagger }A_{j}, \\
\lbrack X,P_{x}] &=&2i\sum\nolimits_{j=1}^{3}\hat{\kappa}_{j}\check{\kappa}%
_{9+j}\left[ 1+\left( q^{2}-1\right) \right] A_{j}^{\dagger }A_{j}, \\
\lbrack Y,P_{y}] &=&-2i\sum\nolimits_{j=1}^{3}\hat{\kappa}_{3+j}\check{\kappa%
}_{12+j}\left[ 1+\left( q^{2}-1\right) \right] A_{j}^{\dagger }A_{j}, \\
\lbrack Z,P_{z}] &=&2i\sum\nolimits_{j=1}^{3}\hat{\kappa}_{6+j}\check{\kappa}%
_{15+j}\left[ 1+\left( q^{2}-1\right) \right] A_{j}^{\dagger }A_{j},
\end{eqnarray}%
\begin{eqnarray}
\lbrack P_{x},P_{y}] &=&-2i\sum\nolimits_{j=1}^{3}\check{\kappa}_{9+j}\check{%
\kappa}_{12+j}\left[ 1+\left( q^{2}-1\right) \right] A_{j}^{\dagger }A_{j},
\label{c4} \\
\lbrack P_{y},P_{z}] &=&2i\sum\nolimits_{j=1}^{3}\check{\kappa}_{12+j}\check{%
\kappa}_{15+j}\left[ 1+\left( q^{2}-1\right) \right] A_{j}^{\dagger }A_{j},
\\
\lbrack X,P_{z}] &=&2i\sum\nolimits_{j=1}^{3}\hat{\kappa}_{j}\check{\kappa}%
_{15+j}\left[ 1+\left( q^{2}-1\right) \right] A_{j}^{\dagger }A_{j}, \\
\lbrack Z,P_{x}] &=&2i\sum\nolimits_{j=1}^{3}\hat{\kappa}_{6+j}\check{\kappa}%
_{9+j}\left[ 1+\left( q^{2}-1\right) \right] A_{j}^{\dagger }A_{j}, \\
\lbrack X,Z] &=&[P_{x},P_{z}]=[X,P_{y}]=[Y,P_{x}]=[Y,P_{z}]=[Z,P_{y}]=0.
\label{c6}
\end{eqnarray}%
Inverting now the relations (\ref{an1})-(\ref{an6}) we find that it is
indeed possible to eliminate entirely the creations and annihilation
operators from these relations. However, this leads to very lengthy
expressions, which we will not present here. Instead we report some special,
albeit still quite general, solutions obtained by setting some of the
constants to zero and imposing further constraints.

\subsection{A particular $\mathcal{PT}_{\pm }$-symmetric solution}

We now make the assumption that $\kappa _{1}=\kappa _{4}=\kappa _{5}=\kappa
_{8}=\kappa _{10}=\kappa _{12}=$ $\kappa _{13}~=\kappa _{14}~=\kappa
_{17}~=\kappa _{18}~=0$. This choice still guarantees that none of the
canonical variables become mutually identical. The consistency with the
direct limit $q\rightarrow 1$ in which we want to recover (\ref{cano})
enforces the constraints 
\begin{equation}
\hat{\kappa}_{2}=\frac{\hbar }{2\check{\kappa}_{11}},~\hat{\kappa}_{3}=\frac{%
\theta _{1}}{2\hat{\kappa}_{6}},~\hat{\kappa}_{9}=-\frac{\theta _{3}}{2\hat{%
\kappa}_{6}},~~\check{\kappa}_{15}=-\frac{\hbar }{2\hat{\kappa}_{6}},~~%
\check{\kappa}_{16}=\frac{\hbar }{2\hat{\kappa}_{7}}.
\end{equation}%
The only non-vanishing commutators we obtain in this case are%
\begin{eqnarray}
\lbrack X,Y] &=&i\theta _{1}+i\frac{q^{2}-1}{q^{2}+1}\frac{\theta _{1}}{%
\hbar }\left( \frac{m\omega }{2\kappa _{6}^{2}}Y^{2}+\frac{2\kappa _{6}^{2}}{%
m\omega }P_{y}^{2}\right) ,  \label{S11} \\
\lbrack Y,Z] &=&i\theta _{3}+i\frac{q^{2}-1}{q^{2}+1}\frac{\theta _{3}}{%
\hbar }\left( \frac{m\omega }{2\kappa _{6}^{2}}Y^{2}+\frac{2\kappa _{6}^{2}}{%
m\omega }P_{y}^{2}\right) ,  \label{S12} \\
\lbrack X,P_{x}] &=&i\hbar +i\frac{q^{2}-1}{q^{2}+1}2m\omega \left( \kappa
_{11}^{2}X^{2}+\frac{1}{4m^{2}\omega ^{2}\kappa _{11}^{2}}P_{x}^{2}+\frac{%
\theta _{1}^{2}\allowbreak \kappa _{11}^{2}}{\hbar ^{2}}P_{y}^{2}+2\frac{%
\theta _{1}\kappa _{11}^{2}}{\hbar }XP_{y}\right) ,~~  \label{S13} \\
\lbrack Y,P_{y}] &=&i\hbar +i\frac{q^{2}-1}{q^{2}+1}2m\omega \left( \frac{1}{%
4\kappa _{6}^{2}}Y^{2}+\frac{\kappa _{6}^{2}}{m^{2}\omega ^{2}}%
P_{y}^{2}\right) ,  \label{S14} \\
\lbrack Z,P_{z}] &=&i\hbar +i\frac{q^{2}-1}{q^{2}+1}2m\omega \left( \frac{1}{%
4\kappa _{7}^{2}}Z^{2}+\frac{\kappa _{7}^{2}}{m^{2}\omega ^{2}}P_{z}^{2}+%
\frac{\theta _{3}^{2}}{4\hbar ^{2}\kappa _{7}^{2}}P_{y}^{2}-\frac{\theta _{3}%
}{2\hbar ^{2}\kappa _{7}^{2}}ZP_{y}\right) .  \label{S15}
\end{eqnarray}%
Notice that we still have the three free parameters $\kappa _{6}$, $\kappa
_{7}$ and $\kappa _{11}$ at our disposal. It is easily verified that the
relations (\ref{S11})-(\ref{S15}) are left invariant under a $\mathcal{PT}%
_{\pm }$-symmetry (\ref{cano}) in the variables $X$, $Y$, $Z$, $P_{x}$, $%
P_{y}$, $P_{z}$.

\subsubsection{Reduced three dimensional solution for $q\rightarrow 1$}

The solution (\ref{S11})-(\ref{S15}) possesses a non-trivial limit leading
to an even simpler set of commutation relations. For this purpose we impose
some additional constraints by setting first $\check{\kappa}_{11}=m\omega 
\hat{\kappa}_{6}$, $\kappa _{7}=1/2\kappa _{6}$, $q=\exp (2\tau \kappa
_{6}^{2})$ and subsequently we take the limit $\kappa _{6}\rightarrow 0$.
The relations (\ref{S11})-(\ref{S15}) then reduce to%
\begin{eqnarray}
\lbrack X,Y] &=&i\theta _{1}\left( 1+\hat{\tau}Y^{2}\right) ,\quad \lbrack
Y,Z]=i\theta _{3}\left( 1+\hat{\tau}Y^{2}\right) ,  \label{S21} \\
\lbrack X,P_{x}] &=&i\hbar \left( 1+\check{\tau}P_{x}^{2}\right) ,\quad
\lbrack Y,P_{y}]=i\hbar \left( 1+\hat{\tau}Y^{2}\right) ,\quad \lbrack
Z,P_{z}]=i\hbar \left( 1+\check{\tau}P_{z}^{2}\right) ,~~  \label{S22}
\end{eqnarray}%
where $\hat{\tau}=\tau m\omega /\hbar $ has the dimension of an inverse
squared length, $\check{\tau}=\tau /(m\omega \hbar )$ has the dimension of
an inverse squared momentum and $\tau ~$is dimensionless. We find a concrete
representation for this algebra in terms of the generators of the standard
three dimensional flat noncommutative space (\ref{cano})%
\begin{equation}
\begin{array}{ll}
X=(1+\check{\tau}p_{_{x_{0}}}^{2})x_{0}+\frac{\theta _{1}}{\hbar }\left( 
\check{\tau}p_{_{x_{0}}}^{2}-\hat{\tau}y_{0}^{2}\right) p_{y_{0}},\qquad  & 
P_{x}=p_{x_{0}}, \\ 
Z=(1+\check{\tau}p_{z_{0}}^{2})z_{0}+\frac{\theta _{3}}{\hbar }\left( \hat{%
\tau}y_{0}^{2}-\check{\tau}p_{z_{0}}^{2}\right) p_{y_{0}}, & P_{z}=p_{z_{0}},
\\ 
P_{y}=(1+\check{\tau}p_{y_{0}}^{2})p_{y_{0}}, & Y=y_{0}.%
\end{array}
\label{rep3d}
\end{equation}%
Evidently the quantities $X$, $Z$ and $P_{y}$ are non-Hermitian in the space
in which the $x_{0}$, $y_{0}$, $z_{0}$, $p_{x_{0}}$, $p_{y_{0}}$, $p_{z_{0}}$
are Hermitian. In order to study concrete models it is very convenient to
carry out a subsequent Bopp-shift of the form $x_{0}\rightarrow x_{s}-\frac{%
\theta _{1}}{\hbar }p_{y_{s}}$, $y_{0}\rightarrow y_{s}$,\ $z_{0}\rightarrow
z_{s}+\frac{\theta _{3}}{\hbar }p_{y_{s}}$, $p_{x_{0}}\rightarrow p_{x_{s}}$%
, $p_{y_{0}}\rightarrow p_{y_{s}}$, $p_{z_{0}}\rightarrow p_{z_{s}}$ and
express the generators in (\ref{rep3d}) in terms of the standard canonical
variables. Since there is no explicit occurrence of $\theta _{2}$, the
representation (\ref{rep3d}) is trivially invariant under $\mathcal{PT}_{\pm
}$ as well as $\mathcal{PT}_{\mathcal{\theta }_{\pm }}$. Taking, however,
the representation (\ref{rep3d}) and in addition $\theta _{2}\neq 0$ this
evidently changes, as by direct computation one of the commutation relations
is altered to $[X,Z]=i\theta _{2}\left( 1+\check{\tau}P_{x}^{2}\right)
\left( 1+\check{\tau}P_{z}^{2}\right) $. Setting furthermore $\theta
_{1}=\theta _{3}$ the representation in (\ref{rep3d}) is also invariant
under the $\mathcal{PT}_{xz}$-symmetry stated in (\ref{xz}).

\subsubsection{Reduction into a decoupled two dimensional plus a one
dimensional space}

The algebra (\ref{S11})-(\ref{S15}) provides a larger three dimensional
setting for a noncommutative two dimensional space decoupled from a standard
one dimensional space. This is achieved by parameterizing $q=\exp (2\tau
\kappa _{6}^{2})$, setting $\check{\kappa}_{11}=m\omega \hat{\kappa}_{6}$, $%
\theta :=\theta _{1}$ and subsequently taking the limit ($\theta _{3}$, $%
\kappa _{6}$)$\rightarrow 0$ reduces the algebra to a two noncommutative
dimensional space in the $X$,$Y$-direction 
\begin{equation}
\begin{array}{lll}
\lbrack X,Y]=i\theta \left( 1+\hat{\tau}Y^{2}\right) ,\qquad & 
[Y,P_{y}]=i\hbar \left( 1+\hat{\tau}Y^{2}\right) ,\qquad & [X,P_{x}]=i\hbar
\left( 1+\check{\tau}P_{x}^{2}\right) ,%
\end{array}
\label{twod}
\end{equation}%
decoupled from a standard one dimensional space in the $Z$-direction%
\begin{equation}
\begin{array}{ll}
\lbrack Z,P_{z}]=i\hbar ,\qquad & [Y,Z]=0.%
\end{array}%
\end{equation}%
As a representation for the algebra (\ref{twod}) in flat noncommutative
space we may simply use (\ref{rep3d}) with the appropriate limit $\theta
_{3}\rightarrow 0$. Carrying out the corresponding Bopp-shift $%
x_{0}\rightarrow x_{s}-\frac{\theta }{\hbar }p_{y_{s}}$, $y_{0}\rightarrow
y_{s}$,\ $p_{x_{0}}\rightarrow p_{x_{s}}$ and $p_{y_{0}}\rightarrow
p_{y_{s}} $ yields the operators%
\begin{equation}
X=x_{s}-\frac{\theta }{\hbar }p_{y_{s}}+\check{\tau}p_{x_{s}}^{2}x_{s}-\hat{%
\tau}\frac{\theta }{\hbar }y_{s}^{2}p_{y_{s}},\quad Y=y_{s},\quad
P_{x}=p_{x_{s}},\quad \text{and\quad }P_{y}=p_{y_{s}}+\hat{\tau}%
y_{s}^{2}p_{y_{s}},  \label{tworep}
\end{equation}%
which are of course still non-Hermitian with regard to the standard inner
product.

\subsubsection{Reduction into three decoupled one dimensional spaces}

We conclude this section by noting that all three directions in the algebra (%
\ref{S11})-(\ref{S15}) can be decoupled, of which one becomes a one
dimensional noncommutative space previously investigated by many authors,
e.g. \cite{Kempf1,AFBB}. It is easy to verify that this scenario is obtained
from (\ref{S11})-(\ref{S15}) when parameterizing $q=\exp (2\tau \kappa
_{11}^{2})$ and subsequently taking the limit $(\theta _{1}$, $\theta _{3}$, 
$\kappa _{11})\rightarrow 0$. The remaining non-vanishing commutators are
then 
\begin{equation}
\lbrack X,P_{x}]=i\hbar \left( 1+\check{\tau}P_{x}^{2}\right) ,\qquad
\lbrack Y,P_{y}]=i\hbar ,\qquad \text{and\qquad }[Z,P_{z}]=i\hbar .
\label{one}
\end{equation}%
Thus all three space directions are decoupled from each other. It is known
that the choices $X=(1+\check{\tau}p_{s}^{2})x_{s}$, $P_{x}=p_{s}$ or $%
X^{\prime }=X^{\dagger }=x_{s}(1+\check{\tau}p_{s}^{2})$, $P_{x}^{\prime
}=p_{s}$ constitute representations for the commutation relations (\ref{one}%
) in the $X$-direction.

\subsection{A particular $\mathcal{PT}_{\mathcal{\protect\theta }_{\pm }}$%
-symmetric solution}

Instead of solving the complicated relations (\ref{sdd})-(\ref{c6}) one may
also start by making directly an Ansatz of a similar form as in (\ref{rep3d}%
) without elaborating on the relation to the $q$-deformed oscillator
algebra. Proceeding in this manner with an Ansatz respecting the $\mathcal{PT%
}_{\mathcal{\theta }_{\pm }}$-symmetry we find for instance the
representation%
\begin{equation}
\begin{array}{ll}
X=x_{0}-\hat{\tau}\frac{\theta _{1}}{\hbar }y_{0}^{2}p_{y_{0}}-\hat{\tau}%
\frac{\theta _{2}}{\hbar }y_{0}^{2}p_{z_{0}}, & P_{x}=p_{x_{0}}, \\ 
Z=z_{0}+\hat{\tau}\frac{\theta _{3}}{\hbar }y_{0}^{2}p_{y_{0}}+\hat{\tau}%
\frac{\theta _{2}}{\hbar }\frac{\theta _{3}}{\theta _{1}}y_{0}^{2}p_{z_{0}},%
\qquad & P_{z}=p_{z_{0}}, \\ 
P_{y}=p_{y_{0}}+\hat{\tau}y_{0}^{2}p_{y_{0}}, & Y=y_{0},%
\end{array}
\label{T2}
\end{equation}%
yielding the closed algebra%
\begin{equation}
\begin{array}{lll}
\lbrack X,Y]=i\theta _{1}\left( 1+\hat{\tau}Y^{2}\right) ,\quad & 
[X,Z]=i\theta _{2}\left( 1+\hat{\tau}Y^{2}\right) ,\quad & [Y,Z]=i\theta
_{3}\left( 1+\hat{\tau}Y^{2}\right) , \\ 
\lbrack X,P_{x}]=i\hbar , & [Y,P_{y}]=i\hbar \left( 1+\hat{\tau}Y^{2}\right)
, & [Z,P_{z}]=i\hbar ,%
\end{array}
\label{t2}
\end{equation}%
with all remaining commutators vanishing. Notice that if we set $\theta
_{1}=-\theta _{3}$ the generators in (\ref{T2}) are also invariant under the 
$\mathcal{PT}_{xz}$-symmetry.

\section{Minimal length, minimal areas and minimal volumes}

Let us now investigate the generalized uncertainty relations associated to
the algebras constructed above. In general, the uncertainties $\Delta A$ and 
$\Delta B$ resulting from a simultaneous measurement of two observables $A$
and $B$ have to obey the inequality%
\begin{equation}
\Delta A\Delta B\geq \frac{1}{2}\left\vert \left\langle [A,B]\right\rangle
_{\rho }\right\vert .  \label{HU}
\end{equation}%
Here $\left\langle .\right\rangle _{\rho }$ denotes the inner product on a
Hilbert space with metric $\rho $ in which the operators $A$ and $B$ are
Hermitian, as discussed in more detail in \cite{Kempf1,AFBB,AFLGFGS,AFLGBB}.
The minimal length $\Delta A_{\min }$, that is the precision up to which the
observable $A$ can be known by giving up all the information on $B$ is then
computed by minimising $\Delta A\Delta B-\frac{1}{2}\left\vert \left\langle
[A,B]\right\rangle _{\rho }\right\vert $ as a function of $\Delta B$. In the
standard scenario, i.e. when $A$ and $B$ commute up to a constant, the
result is therefore usually zero. This outcome changes when the commutator $%
[A,B]$ involves higher powers of $\Delta B$, in which case we encounter the
interesting scenario of non-vanishing $\Delta A_{\min }$. We now investigate
some of the solutions presented above. Depending now on the question we ask,
i.e. which quantities we attempt to measure, the minimal uncertainties for
some specific operators turn out to be different.

\subsection{A three dimensional noncommutative space giving rise to minimal
areas}

We start with our simplest three dimensional solution, that is the algebra (%
\ref{S21})-(\ref{S22}). If we just want to measure the position of the
particle on such a space independently of its momentum we only have to
investigate the relations (\ref{S21}). Taking $\tau >0$ and following the
logic of \cite{AFBB,AFLGFGS,AFLGBB}, we obtain from (\ref{S21}) for a
simultaneous measurement of all space coordinates non-vanishing minimal
length in two directions 
\begin{equation}
\Delta X_{\min }=\left\vert \theta _{1}\right\vert \sqrt{\hat{\tau}}\sqrt{1+%
\hat{\tau}\left\langle Y\right\rangle _{\rho }^{2}},\quad \Delta Y_{\min
}=0,\quad \text{and\quad }\Delta Z_{\min }=\left\vert \theta _{3}\right\vert 
\sqrt{\hat{\tau}}\sqrt{1+\hat{\tau}\left\langle Y\right\rangle _{\rho }^{2}}.
\label{xmin1}
\end{equation}%
Thus any measurement of space will involve an unavoidable uncertainty of an
area $A$ of size $\Delta A_{0}=4\hat{\tau}\left\vert \theta _{1}\theta
_{3}\right\vert $ in the $XZ$-plane and no uncertainty in the $Y$-direction.
Changing our question and attempt to measure instead all coordinates and all
components of the momenta, we need to analyze the entire set of relations (%
\ref{S21})-(\ref{S22}). The analysis of the equations (\ref{S22}) alone
yields%
\begin{eqnarray}
\Delta X_{\min } &=&\hbar \sqrt{\check{\tau}}\sqrt{1+\hat{\tau}\left\langle
Y\right\rangle _{\rho }^{2}},\quad \Delta Y_{\min }=0,\quad \text{and\quad }%
\Delta Z_{\min }=\hbar \sqrt{\check{\tau}}\sqrt{1+\hat{\tau}\left\langle
Y\right\rangle _{\rho }^{2}},~~~  \label{xmin3} \\
\Delta \left( P_{x}\right) _{\min } &=&0,\quad \ \ \ \ \ \ \ \ \ \ \Delta
\left( P_{y}\right) _{\min }=\hbar \sqrt{\hat{\tau}}\sqrt{1+\hat{\tau}%
\left\langle Y\right\rangle _{\rho }^{2}},\quad \text{and\quad }\Delta
\left( P_{z}\right) _{\min }=0.~~  \label{xmin2}
\end{eqnarray}%
Thus, depending now on whether $\left\vert \theta _{1}\right\vert $, $%
\left\vert \theta _{3}\right\vert <1$ or $\left\vert \theta _{1}\right\vert $%
, $\left\vert \theta _{3}\right\vert >1$ the uncertainties in (\ref{xmin1})
or (\ref{xmin3}) will be smaller, respectively. For any type of measurement
the region of uncertainty will be an area.

\subsection{A three dimensional noncommutative space giving rise to minimal
volumes}

Let us now analyze our solution (\ref{S21})-(\ref{S22}) before taking the
limit $q\rightarrow 0$. We compute the uncertainties with regard to a
measurement of all components of the coordinates and all components of the
momenta. Since now the quantities are all coupled, in the sense that we do
not have any nontrivial subalgebra, we will encounter uncertainties for all
of them and observe a different type of behaviour as indicated in the
previous subsection. Starting with a simultaneous $Y$,$P_{y}$-measurement we
compute from (\ref{HU}) with (\ref{S14}) the uncertainties%
\begin{eqnarray}
\Delta Y_{\min } &=&\left\vert \hat{\kappa}_{6}\right\vert \sqrt{\frac{1}{2}%
(q^{2}-q^{-2})+(q-q^{-1})^{2}\left( \frac{1}{4\hat{\kappa}_{6}^{2}}%
\left\langle Y\right\rangle _{\rho }^{2}+\frac{\hat{\kappa}_{6}^{2}}{\hbar
^{2}}\left\langle P_{y}^{2}\right\rangle _{\rho }\right) },  \label{min1} \\
\Delta \left( P_{y}\right) _{\min } &=&\frac{\hbar }{2\left\vert \hat{\kappa}%
_{6}\right\vert }\sqrt{\frac{1}{2}(q^{2}-q^{-2})+(q-q^{-1})^{2}\left( \frac{1%
}{4\hat{\kappa}_{6}^{2}}\left\langle Y\right\rangle _{\rho }^{2}+\frac{\hat{%
\kappa}_{6}^{2}}{\hbar ^{2}}\left\langle P_{y}^{2}\right\rangle _{\rho
}\right) },  \label{min2}
\end{eqnarray}%
under the assumption that $q>1$. The absolute minimal uncertainties
resulting from these expressions are therefore%
\begin{equation}
\Delta Y_{0}=\frac{\left\vert \hat{\kappa}_{6}\right\vert }{\sqrt{2}}\sqrt{%
q^{2}-q^{-2}},\qquad \text{and\qquad }\Delta \left( P_{y}\right) _{0}=\frac{%
\hbar }{2\sqrt{2}\left\vert \hat{\kappa}_{6}\right\vert }\sqrt{q^{2}-q^{-2}}.
\label{P0}
\end{equation}%
Next we carry out a simultaneous $X$,$Y$-measurement and a $Y$,$Z$%
-measurement by employing (\ref{S11}) and (\ref{S12}), respectively. We find
the minimal lengths%
\begin{eqnarray}
\Delta X_{\min } &=&\left\vert \frac{\theta _{1}}{\hat{\kappa}_{6}}%
\right\vert \sqrt{\frac{1}{2}\frac{q-q^{-1}}{q+q^{-1}}+\left[ \frac{q-q^{-1}%
}{q+q^{-1}}\right] ^{2}\left[ \frac{1}{4\hat{\kappa}_{6}^{2}}\left\langle
Y\right\rangle _{\rho }^{2}+\frac{\hat{\kappa}_{6}^{2}}{\hbar ^{2}}\left(
\left\langle P_{y}^{2}\right\rangle _{\rho }+\Delta \left( P_{y}\right)
_{0}^{2}\right) \right] }, \\
\Delta Z_{\min } &=&\left\vert \frac{\theta _{3}}{\hat{\kappa}_{6}}%
\right\vert \sqrt{\frac{1}{2}\frac{q-q^{-1}}{q+q^{-1}}+\left[ \frac{q-q^{-1}%
}{q+q^{-1}}\right] ^{2}\left[ \frac{1}{4\hat{\kappa}_{6}^{2}}\left\langle
Y\right\rangle _{\rho }^{2}+\frac{\hat{\kappa}_{6}^{2}}{\hbar ^{2}}\left(
\left\langle P_{y}^{2}\right\rangle _{\rho }+\Delta \left( P_{y}\right)
_{0}^{2}\right) \right] }.~
\end{eqnarray}%
There is no minimal length in the $Y$-direction resulting from these
relations. Using the expression for $\Delta \left( P_{y}\right) _{0}$ from (%
\ref{P0}), the absolute minimal values for these uncertainties are%
\begin{equation}
\Delta X_{0}=\frac{1}{2\sqrt{2}}\left\vert \frac{\theta _{1}}{\hat{\kappa}%
_{6}}\right\vert \sqrt{q^{2}-q^{-2}},\qquad \text{and\qquad }\Delta Z_{0}=%
\frac{1}{2\sqrt{2}}\left\vert \frac{\theta _{3}}{\hat{\kappa}_{6}}%
\right\vert \sqrt{q^{2}-q^{-2}}.
\end{equation}%
Thus a measurement of the position in space will be accompanied by an
uncertainty volume $V$ of the size 
\begin{equation}
\Delta V_{0}=\frac{1}{\sqrt{2}}\left\vert \frac{\theta _{1}\theta _{3}}{\hat{%
\kappa}_{6}}\right\vert \left( q^{2}-q^{-2}\right) ^{3/2}.
\end{equation}%
The evaluation for the simultaneous $X$,$P_{x}$ and $Z$,$P_{z}$-measurements
are slightly more complicated due to the occurrence of the $XP_{y}$ and $%
ZP_{y}$ terms in (\ref{S13}) and (\ref{S15}), respectively. We proceed
similarly as before and make also use of the well known inequalities $%
\left\vert A+B\right\vert \geq \left\vert A\right\vert -\left\vert
B\right\vert $ and $\left\vert \left\langle AB\right\rangle \right\vert \leq
\Delta A\Delta B+\left\vert \left\langle A\right\rangle \left\langle
B\right\rangle \right\vert $. We report here only the final result of the
absolute minimal values%
\begin{equation}
\Delta \left( P_{i}\right) _{0}=\frac{\gamma _{i}\Delta \left( P_{y}\right)
_{0}-\sqrt{\beta _{i}\left[ \alpha _{i}\gamma _{i}^{2}\Delta \left(
P_{y}\right) _{0}^{2}+\lambda _{i}(1-4\alpha _{i}\beta _{i})\right] }}{%
4\alpha _{i}\beta _{i}-1}\quad ~~~\text{for }i=x,z,
\end{equation}%
with 
\begin{equation}
\begin{array}{llll}
\alpha _{x}=\alpha _{2},~~ & \beta _{x}=\alpha _{11},~~ & \gamma _{x}=\frac{%
2\left\vert \theta _{1}\right\vert }{\hbar }\alpha _{11},~~ & \lambda _{x}=%
\frac{\hbar }{2}+\alpha _{11}\frac{\theta _{1}^{2}}{\hbar ^{2}}\Delta \left(
P_{y}\right) _{0}^{2}, \\ 
\alpha _{z}=\alpha _{7},~~ & \beta _{z}=\alpha _{16},~~ & \gamma _{z}=\frac{%
\left\vert \theta _{3}\right\vert \hbar }{2}\alpha _{16},~~ & \lambda _{z}=%
\frac{\hbar }{2}+\alpha _{16}\theta _{3}^{2}\Delta \left( P_{y}\right)
_{0}^{2},%
\end{array}%
\end{equation}%
where $\alpha _{i}=$\ $\hat{\kappa}_{i}^{2}(q-q^{-1})/(q+q^{-1})\hbar $ for $%
i=2,7$ and $\alpha _{i}=$\ $\check{\kappa}_{i}^{2}(q-q^{-1})/(q+q^{-1})\hbar 
$ for $i=11,16$. Further restrictions do not emerge.

By similar reasoning one finds non-vanishing $\Delta X_{\min }$, $\Delta
Z_{\min }$ and $\Delta P_{y\min }$ for the $\mathcal{PT}_{xz}$-invariant
algebra (\ref{t2}).

\section{Models on $\mathcal{PT}$-symmetric noncommutative spaces}

\subsection{The one dimensional harmonic oscillator on a noncommutative space%
}

We commence with the one-dimensional harmonic oscillator on the $\mathcal{PT}%
_{\pm }$-symmetric noncommutative space described by (\ref{one}). The
corresponding Hamiltonian 
\begin{equation}
H_{ncho}^{1D}=\frac{P^{2}}{2m}+\frac{m\omega ^{2}}{2}X^{2}=H_{ho}^{1D}+\frac{%
m\omega ^{2}}{2}\left( \check{\tau}p_{s}^{2}x_{s}^{2}+\check{\tau}%
x_{s}p_{s}^{2}x_{s}+\check{\tau}^{2}p_{s}^{2}x_{s}p_{s}^{2}x_{s}\right)
=H_{ho}^{1D}+H_{nc}^{1D},  \label{HD1}
\end{equation}%
is evidently non-Hermitian with regard to the standard inner product.
However, it is $\mathcal{PT}_{\pm }$-symmetric, such that it might
constitutes a well-defined self-consistent description of a physical system.
The associated Schr\"{o}dinger equation $H_{ncho}^{1D}\psi =E\psi $ is most
conveniently solved in $p$-space, i.e. with $x_{s}=i\hbar \partial _{p_{s}}$
it reads%
\begin{equation}
\frac{m\omega ^{2}\hbar ^{2}}{2}(1+\check{\tau}p_{s}^{2})^{2}\psi ^{\prime
\prime }+\tau \omega \hbar p_{s}(1+\check{\tau}p_{s}^{2})\psi ^{\prime
}+\left( E-\frac{p_{s}^{2}}{2m}\right) \psi =0.  \label{diff}
\end{equation}%
Using the transformation%
\begin{equation}
\mu =\frac{\sqrt{1+2Em\check{\tau}}}{\tau },\qquad \nu =\frac{\sqrt{4+\tau
^{2}}}{2\tau }-\frac{1}{2}\quad \text{and\quad }z=ip_{s}\sqrt{\check{\tau}},
\label{munu}
\end{equation}%
we convert (\ref{diff}) into 
\begin{equation}
(1-z^{2})\psi ^{\prime \prime }-2z\psi ^{\prime }+\left[ \nu (\nu +1)-\frac{%
\mu ^{2}}{1-z^{2}}\right] \psi =0,  \label{al}
\end{equation}%
which is the standard differential equation for the associated Legendre
polynomials $P_{\nu }^{\mu }(z)$ and $Q_{\nu }^{\mu }(z)$ admitting the
general solution%
\begin{equation}
\psi (z)=c_{1}P_{\nu }^{\mu }(z)+c_{2}Q_{\nu }^{\mu }(z).
\end{equation}%
Seeking asymptotically vanishing solutions gives rise to the quantization
condition $\mu +\nu =-n-1$ with $n\in \mathbb{N}$. With (\ref{munu}) it
follows therefore that the eigenenergies becomes%
\begin{equation}
E_{n}=\omega \hbar \left( \frac{1}{2}+n\right) \sqrt{1+\frac{\tau ^{2}}{4}}%
+\tau \frac{\omega \hbar }{4}(1+2n+2n^{2})\quad \quad \text{for }n\in 
\mathbb{N}_{0}.  \label{En}
\end{equation}%
The expression agrees with the one found in \cite{Kempf1}. The polynomial $%
Q_{\nu }^{\mu }(z)$ is not defined for these values, such that $c_{2}=0$ and 
$P_{\nu }^{\mu }(z)$ reduces to 
\begin{equation}
\psi _{2n-i}(z)=c_{1}\sum_{k=i}^{2n-i}\frac{1}{k!}\left[ \prod\limits_{l=i}^{%
\frac{k+i-2}{2}}2(n-l)(2n+2\nu -2l+1)\right] z^{k}\frac{(z^{2}-1)^{\frac{\nu
-2n-1+i}{2}}}{(-1)^{1+i}(1-z^{2})^{\nu }},
\end{equation}%
with $i=0,1$. Clearly the $\psi _{2n-i}(z)$ vanish for $\left\vert
z\right\vert \rightarrow \infty $ if $\nu >-1$, which is always guaranteed
for $\tau m\omega >0$. The Dyson map $\eta $ which adjointly maps $%
H_{ncho}^{1D}$ to a Hermitian operator was easily found \cite{Kempf1,AFBB}
to be $\eta =\left( 1+\tau P_{x}^{2}\right) ^{-1/2}$. In addition, we note
that the solutions are square integrable $\psi _{2n-i}(z)\in L^{2}(i\mathbb{R%
})$ on $\left\langle \cdot \right\vert \left. \eta ^{2}\cdot \right\rangle $
and form an orthonormal basis.

An exact treatment for models in the higher dimensions is more difficult,
but we may resort to perturbation theory to obtain some useful insight on
the solutions. As a quality gauge we compare here the exact solution against
perturbation theory around the standard Fock space harmonic oscillator
solution with normalized eigenstates%
\begin{equation}
\left\vert n\right\rangle =\frac{(a_{x_{s}}^{\dagger })^{n}}{\sqrt{n!}}%
\left\vert 0\right\rangle ,\quad a_{x_{s}}\left\vert 0\right\rangle =0,\quad
a_{x_{s}}^{\dagger }\left\vert n\right\rangle =\sqrt{n+1}\left\vert
n+1\right\rangle ,\quad a_{x_{s}}\left\vert n\right\rangle =\sqrt{n}%
\left\vert n-1\right\rangle .
\end{equation}%
A straightforward, albeit lengthy, computation yields the following
corrections to the harmonic oscillator energy $E_{n}^{(0)}=\omega \hbar
\left( n+\frac{1}{2}\right) $ for the eigenenergies of $H_{ncho}^{1D}$ 
\begin{equation}
E_{n}^{(p)}=E_{n}^{(0)}+E_{n}^{(1)}+E_{n}^{(2)}+\mathcal{O}(\tau ^{3})
\end{equation}%
with%
\begin{eqnarray}
E_{n}^{(1)} &=&\left\langle n\right\vert H_{nc}^{1D}\left\vert
n\right\rangle =\frac{\tau \omega \hbar }{4}\left( 1+2n+2n^{2}\right) +\frac{%
\tau ^{2}\omega \hbar }{16}\left( 3+8n+6n^{2}+4n^{3}\right) ,~~~~ \\
E_{n}^{(2)} &=&\sum_{p\neq n}\frac{\left\langle n\right\vert
H_{nc}^{1D}\left\vert p\right\rangle \left\langle p\right\vert
H_{nc}^{1D}\left\vert n\right\rangle }{E_{n}^{(0)}-E_{p}^{(0)}}=-\frac{1}{8}%
\tau ^{2}\omega \hbar \left( 1+3n+3n^{2}+2n^{3}\right) +\mathcal{O}(\tau
^{3}).~~
\end{eqnarray}%
As it should be, the expression for $E_{n}$ in (\ref{En}) when expanded up
to order $\tau ^{3}$ coincides precisely with $E_{n}^{(p)}$. We further note
that also in a perturbative treatment the eigenenergies are strictly
positive.

The validity of these expansions is governed by the well-known sufficient
conditions for the applicability of the Rayleigh-Schr\"{o}dinger
perturbation theory to a Hamiltonian of the form $H=H_{0}+H_{1}$ around the
solutions of $H_{0}\left\vert n\right\rangle =E_{n}^{(0)}\left\vert
n\right\rangle $%
\begin{equation}
\left\vert \frac{\left\langle p\right\vert H_{1}\left\vert n\right\rangle }{%
E_{n}^{(0)}-E_{p}^{(0)}}\right\vert \ll 1\qquad \text{for all }p\neq n.
\end{equation}%
This is guaranteed for (\ref{HD1}) when $\tau ^{2}\ll 32/(2n+13)\sqrt{%
(4+n)(3+n)(2+n)(1+n)}$, such that perturbation theory will break down for
large values of $n$.

\subsection{The two dimensional harmonic oscillator on a noncommutative space%
}

Next we consider the two-dimensional harmonic oscillator on the $\mathcal{PT}%
_{\pm }$-symmetric noncommutative space described by the algebra (\ref{twod}%
). Using the representation (\ref{tworep}) for this algebra, the
corresponding Hamiltonian reads 
\begin{eqnarray}
H_{ncho}^{2D} &=&\frac{1}{2m}(P_{x}^{2}+P_{y}^{2})+\frac{m\omega ^{2}}{2}%
(X^{2}+Y^{2})  \label{2dho} \\
&=&H_{fncho}^{2D}+\frac{\tau \omega }{2\hbar }\left[
\{p_{x_{0}}^{2}x_{0},x_{0}\}+\{y_{0}^{2}p_{y_{0}},p_{y_{0}}\}+\frac{\theta }{%
\hbar }\{p_{x_{0}}^{2}p_{y_{0}},x_{0}\}-\frac{m^{2}\omega ^{2}\theta }{\hbar 
}\{y_{0}^{2}p_{y_{0}},x_{0}\}\right]   \notag \\
&&+\frac{\tau ^{2}\omega ^{2}m}{2\hbar ^{2}}\left[ \left\{
y_{0}^{2}p_{y_{0}},(1+\Omega )y_{0}^{2}p_{y_{0}}-\frac{\theta
p_{x_{0}}^{2}x_{0}}{\hbar }-\frac{\theta ^{2}p_{x_{0}}^{2}p_{y_{0}}}{\hbar
^{2}}\right\} +\left( \frac{p_{x_{0}}^{2}x_{0}}{m\omega }+\frac{\theta
p_{x_{0}}^{2}p_{y_{0}}}{m\omega \hbar }\right) ^{2}\right]   \notag
\end{eqnarray}%
where we used the standard notation for the anti-commutator $\left\{
A,B\right\} =AB+BA$. Once again this Hamiltonian is non-Hermitian with
regard to the inner product on the flat noncommutative space, but it
respects a $\mathcal{PT}_{\pm }$-symmetry. In order to be able to perturb
around the standard harmonic oscillator solution we still need to convert
flat noncommutative space into the canonical variable $x_{s}$, $y_{s}$, $%
p_{x_{s}}$ and $p_{y_{s}}$. Thus when using the representation (\ref{tworep}%
) this Hamiltonian is converted into 
\begin{eqnarray*}
H_{ncho}^{2D} &=&H_{ho}^{2D}+\frac{m\theta ^{2}\omega ^{2}}{2\hbar ^{2}}%
p_{y_{s}}^{2}-\frac{m\theta \omega ^{2}}{2\hbar }\{x_{s},p_{y_{s}}\}+\frac{%
\tau }{2}\left[ m\omega ^{2}\{p_{x_{s}}^{2}x_{s},x_{s}\}\right.  \\
&&+\left. \left( \frac{1}{m}+\frac{m\theta ^{2}\omega ^{2}}{\hbar ^{2}}%
\right) \{y_{s}^{2}p_{y_{s}},p_{y_{s}}\}-\frac{m\theta \omega ^{2}}{\hbar }%
\left( \{y_{s}^{2}p_{y_{s}},x_{s}\}+\{p_{x_{s}}^{2}x_{s},p_{y_{s}}\}\right) %
\right]  \\
&&+\frac{\tau ^{2}}{2}\left[ \left( \frac{1}{m}+\frac{m\theta ^{2}\omega ^{2}%
}{\hbar ^{2}}\right) \left( y_{s}^{2}p_{y_{s}}\right) ^{2}+m\omega
^{2}\left( p_{x_{s}}^{2}x_{s}\right) ^{2}-\frac{m\theta \omega ^{2}}{\hbar }%
\{y_{s}^{2}p_{y_{s}},p_{x_{s}}^{2}x_{s}\}\right] , \\
&=&H_{ho}^{2D}(x_{s},y_{s},p_{x_{s}},p_{y_{s}})+H_{nc}^{2D}(x_{s},y_{s},p_{x_{s}},p_{y_{s}}).
\end{eqnarray*}

In this formulation we may now proceed as in the previous subsection and
expand perturbatively around the standard two dimensional Fock space
harmonic oscillator solution with normalized eigenstates%
\begin{eqnarray}
\left\vert n_{1}n_{2}\right\rangle &=&\frac{(a_{1}^{\dagger
})^{n_{1}}(a_{2}^{\dagger })^{n_{2}}}{\sqrt{n_{1}!n_{2}!}}\left\vert
00\right\rangle ,\quad a_{i}^{\dagger }\left\vert n_{1}n_{2}\right\rangle =%
\sqrt{n_{i}+1}\left\vert (n_{1}+\delta _{i1})(n_{2}+\delta
_{i2})\right\rangle ,\quad \\
a_{i}\left\vert 00\right\rangle &=&0,~~~~~\qquad \qquad \ \ \ \ \ \ \ \
a_{i}\left\vert n_{1}n_{2}\right\rangle =\sqrt{n_{i}}\left\vert
(n_{1}-\delta _{i1})(n_{2}-\delta _{i2})\right\rangle ,
\end{eqnarray}%
for $i=1,2$. The energy eigenvalues for the Hamiltonian $H_{ncho}^{2D}$ then
result to 
\begin{eqnarray}
E_{nl}^{p} &=&E_{nl}^{(0)}+E_{nl}^{(1)}+E_{nl}^{(2)}+\mathcal{O}(\tau ^{2})
\\
&=&\hbar \omega (n+l+1)+\left\langle nl\right\vert H_{nc}^{2D}\left\vert
nl\right\rangle +\sum_{p,q\neq n+l=p+q}\frac{\left\langle nl\right\vert
H_{nc}^{2D}\left\vert pq\right\rangle \left\langle pq\right\vert
H_{nc}^{2D}\left\vert nl\right\rangle }{E_{nl}^{(0)}-E_{pq}^{(0)}}+\mathcal{O%
}(\tau ^{2})  \notag \\
&=&E_{nl}^{(0)}+\frac{\Omega \omega \hbar }{8}\left[ (3+n+5l)-\Omega (l+%
\frac{1}{2})\right] +\frac{\tau }{2}\omega \hbar \left[ 1+n+n^{2}+l+l^{2}%
\right.  \notag \\
&&\left. \frac{\Omega }{4}\left( 4+3n+n^{2}+7l+4nl+5l^{2}\right) \right] +%
\mathcal{O}(\tau ^{2}),  \notag
\end{eqnarray}%
where we introduced the dimensionless quantity $\Omega _{i}=m^{2}\theta
_{i}^{2}\omega ^{2}/\hbar ^{2}$. Notice that unlike as in the one
dimensional case the perturbation beyond $H_{ho}^{2D}$ also involves terms
of order $\mathcal{O}(\tau ^{0})$, such that we need to compute also $%
E_{nl}^{(2)}$ to achieve a precision of first order in $\tau $. We also
notice that the energy $E_{nl}^{p}$ is only bounded from below for $\Omega
<5 $. The minus sign is an indication that we will encounter exceptional
points \cite{CarlExPoint} and broken $\mathcal{PT}$-symmetry in some
parameter range.\ 

\subsection{The three dimensional harmonic oscillator on a noncommutative
space}

Let us finally consider the three-dimensional harmonic oscillator on the
noncommutative space described by the algebra (\ref{S11}) and (\ref{S12}).
Using the representation (\ref{rep3d}) together with a subsequent
Bopp-shift, the corresponding Hamiltonian can be expressed in terms of the
standard canonical coordinates%
\begin{eqnarray}
H_{ncho}^{3D} &=&\frac{1}{2m}(P_{x}^{2}+P_{y}^{2}+P_{z}^{2})+\frac{m\omega
^{2}}{2}(X^{2}+Y^{2}+Z^{2})=H_{ho}^{3D}+H_{nc}^{3D} \\
&=&H_{ho}^{3D}+\frac{m\omega ^{2}}{2\hbar }\left[ \theta
_{3}\{p_{y_{s}},z_{s}\}-\theta _{1}\{x_{s},p_{y_{s}}\}+\frac{\theta
_{1}^{2}+\theta _{3}^{2}}{\hbar }p_{y_{s}}^{2}\right]   \notag \\
&&+\tau \frac{\omega }{2\hbar }\left[ \{p_{x_{s}}^{2}x_{s},x_{s}\}+%
\{p_{z_{s}}^{2}z_{s},z_{s}\}+\left( 1+\Omega _{1}+\Omega _{3}\right)
\{y_{s}^{2}p_{y_{s}},p_{y_{s}}\}\right.   \notag \\
&&-\left. \frac{\theta _{1}}{\hbar }\left( m^{2}\omega
^{2}\{y_{s}^{2}p_{y_{s}},x_{s}\}+\{p_{x_{s}}^{2}x_{s},p_{y_{s}}\}\right) +%
\frac{\theta _{3}}{\hbar }\left( m^{2}\omega
^{2}\{y_{s}^{2}p_{y_{s}},z_{s}\}+\{p_{z_{s}}^{2}z_{s},p_{y_{s}}\}\right) %
\right]   \notag \\
&&+\tau ^{2}\frac{1}{2m\hbar ^{2}}\left[
p_{x_{s}}^{2}x_{s}p_{x_{s}}^{2}x_{s}+p_{z_{s}}^{2}z_{s}p_{z_{s}}^{2}\text{ }%
z_{s}+m^{2}\omega ^{2}\left( 1+\Omega _{1}+\Omega _{3}\right)
y_{s}^{2}p_{y_{s}}y_{s}^{2}p_{y_{s}}\right.   \notag \\
&&\left. +\frac{\theta _{3}}{\hbar }m^{2}\omega
^{2}\{p_{z_{s}}^{2}z_{s},y_{s}^{2}p_{y_{s}}\}-\frac{\theta _{1}}{\hbar }%
m^{2}\omega ^{2}\{p_{x_{s}}^{2}x_{s},y_{s}^{2}p_{y_{s}}\}\right]   \notag
\end{eqnarray}%
We expand now around the standard three dimensional Fock space harmonic
oscillator solution with normalized eigenstates%
\begin{eqnarray}
\left\vert n_{1}n_{2}n_{3}\right\rangle  &=&\prod\nolimits_{i=1}^{3}\frac{%
(a_{i}^{\dagger })^{n_{i}}}{\sqrt{n_{i}!}}\left\vert 000\right\rangle ,\quad
a_{i}^{\dagger }\left\vert n_{1}n_{2}n_{3}\right\rangle =\sqrt{n_{i}+1}%
\left\vert \prod\nolimits_{j=1}^{3}(n_{j}+\delta _{ij})\right\rangle ,\quad 
\\
a_{i}\left\vert 000\right\rangle  &=&0,~~~~~\qquad \qquad \ \ \ \ \ \ \ \
a_{i}\left\vert n_{1}n_{2}n_{3}\right\rangle =\sqrt{n_{i}}\left\vert
\prod\nolimits_{j=1}^{3}(n_{j}-\delta _{ij})\right\rangle .
\end{eqnarray}%
for $i=1,2,3$ and compute the energy eigenvalues for $H_{ncho}^{3D}$ to%
\begin{eqnarray}
E_{nlr}^{(p)} &=&E_{nlr}^{(0)}+E_{nlr}^{(1)}+E_{nlr}^{(2)}+\mathcal{O}(\tau
^{2}) \\
&=&E_{nl}^{(0)}+\left\langle nlr\right\vert H_{nc}^{3D}\left\vert
nlr\right\rangle +\sum_{s,p,q\neq n+l+r=p+q+s}\frac{\left\langle
nlr\right\vert H_{nc}^{3D}\left\vert pqs\right\rangle \left\langle
pqs\right\vert H_{nc}^{3D}\left\vert nlr\right\rangle }{%
E_{nlr}^{(0)}-E_{pqr}^{(0)}}+\mathcal{O}(\tau ^{2})  \notag \\
&=&\omega \hbar \left[ \frac{3}{2}+n+l+r+\frac{1}{8}\left( \Omega
_{1}+\Omega _{3}\right) \left( 3+5l\right) -\frac{1}{16}(2l+1)\left( \Omega
_{1}+\Omega _{3}\right) {}^{2}+\frac{1}{8}\left( n\Omega _{1}+r\Omega
_{3}\right) \right.   \notag \\
&&+\frac{\tau }{2}\left( n^{2}+n+l^{2}+r^{2}+l+r+\frac{1}{4}\left(
n^{2}+4ln+3n+5l^{2}+7l+4\right) \Omega _{1}\right.   \notag \\
&&+\left. \left. \frac{1}{4}\left( 5l^{2}+4rl+7l+r^{2}+3r+4\right) \Omega
_{3}+\frac{3}{2}\right) \right] .  \notag
\end{eqnarray}%
As in the two dimensional case we encounter negative terms in this
expression, thus indicating that broken $\mathcal{PT}$-symmetry will be
broken in some parameter range.\ 

\section{Conclusions}

Contrary to some claims in the literature \cite{Giri:2008iq}, we have
demonstrated that it is indeed possible to implement $\mathcal{PT}$-symmetry
on noncommutative spaces while keeping the noncommutative constants real.
Starting from a generic Ansatz for the canonical variables obeying a
q-deformed oscillator algebra, we employed $\mathcal{PT}$-symmetry to limit
the amount of free parameters. The relations (\ref{sdd})-(\ref{c6})
resulting from this Ansatz turned out to be solvable. A specific $\mathcal{PT%
}_{\pm }$-symmetric solution was presented in (\ref{S11})-(\ref{S15}).
Clearly there exist more solutions with different kinds of properties. We
constructed an explicit representation for the algebra obtained in the
nontrivial limit $q\rightarrow 1$ in terms of the generators of a flat
noncommutative space. With regard to the standard inner product for this
space, the operators are non-Hermitian. We computed the minimal length and
momenta resulting from the generalized uncertainty relations, which overall
give rise to minimal areas or minimal volumes in phase space.

Despite being non-Hermitian, due to the built-in $\mathcal{PT}$-symmetry any
model formulated in terms of these variables is a candidate for a
self-consistent theory with real eigenvalue spectrum. We have studied the
harmonic oscillator on these spaces in one, two and three dimensions. The
perturbative computation of the energy eigenvalues indicates that there
exists a parameter regime for which the $\mathcal{PT}$-symmetry is broken.
It would be interesting to investigate this further and determine when this
transition precisely occurs. The eigenvalues will also be useful in further
investigations \cite{SDAFprep} allowing for the construction of coherent
states related to the algebras presented in section 4.

Obviously there are many more solutions to (\ref{sdd})-(\ref{c6}), which
might be studied in their own right together with models formulated on them.
Minor modifications would also allow to investigate the occurrence of upper
bounds, i.e. maximal length and momenta \cite{Nozari:2012gd}, giving rise to
a second scale in special relativity, so-called double special relativity 
\cite{Amelino} constituting a possibility to explain the cosmic-ray paradox 
\cite{Cosmic}.\medskip

\noindent \textbf{Acknowledgments:} SD is supported by a City University
Research Fellowship. LG is supported by the high energy section of the ICTP.


\end{document}